# Measuring the position of the center of the Sun at the Clementine Gnomon of Santa Maria degli Angeli in Rome

Costantino Sigismondi, Galileo Ferraris Institute and ICRA International Center for Relativistic Astrophysics, Rome.

**Abstract:**

The Clementine Gnomon in the Basilica of Santa Maria degli Angeli in Rome has been realized in 1702 with the aim to measure the variation of the obliquity of the Earth axis along the forthcoming centuries. Since then the church and the instrument undergone several restorations and the original conditions of the pinhole changed.

The measurements of the position of the image in the days before and of the 2011 winter solstice with respect to the original markers compared with the ephemerides gives us the North-South correction for the position of the pinhole to be restored.

**Introduction:**

The Clementine Gnomon is a solar meridian telescope dedicated to solar astrometry operating as a giant pinhole dark camera, being the basilica of Santa Maria degli Angeli the dark room. This instrument built in 1701-1702 by the will of pope Clement XI by Francesco Bianchini (1662-1729) gives solar images free from distortions, excepted atmospheric refraction, because the pinhole is optics less. Similar historical instruments are in Florence (Duomo, by Toscanelli and Ximenes), Bologna (San Petronio, by Cassini), Milan (Duomo, by De Cesaris) and Palermo (Cathedral, by Piazzi).[1] The azimut of the Clementine Gnomon has been referenced with respect to the celestial North pole, and it is 4'28.8"±0.6",[2, 4].

**Local deviations from a perfect line and evaluation of ΔUT1:**

Also the local deviations from a perfect line are known with accuracy better than 0.5 mm, by using a LASER beam.
Due to the velocity of the drift of the solar image up to 3 mm/s, these local deviations add time scatters up to 2 s to the above mentioned global eastward deviation, which attains a maximum delay of ~18 s at the winter solstice, with respect to the modern ephemerides time.

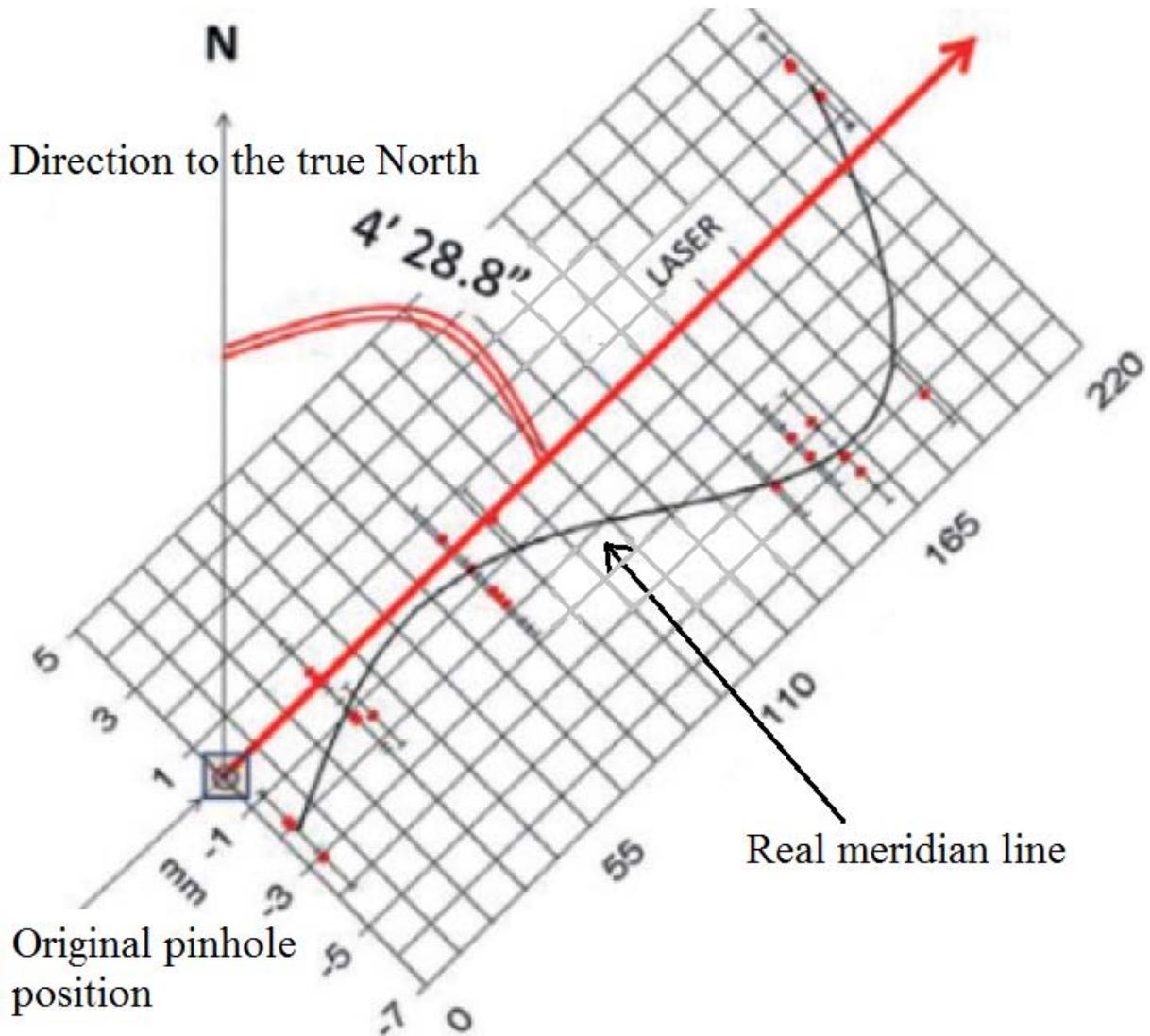

**Fig. 1** Local deviation of the Clementine Line from the LASER line, and global deviation from the true North.

With these calibration data we used the Gnomon to measure the delay of the solar meridian transit with respect to the time calculated by the ephemerides. Since the ephemerides are computed with a constant Earth rotation rate, this difference is a direct measurement of ΔUT1. The growth of this astronomical parameter is compensated by the insertion of a leap second ad the end of the year in order to keep the Universal Time close to astronomical phenomena within less than a whole second.[1]

On December 31, 2008 at 23:59:59 there is one of those leap seconds leading to 23:59:60 before the new year's midnight 00:00:00, being ΔUT1~0.7 s at that date. ΔUT1 has been measured in that occasion with an accuracy of ±0.3 s. [2]

The reason of such limit in time accuracy is due to the effect of the atmospheric seeing.

---

[1] see http://hpiers.obspm.fr/iers/bul/bulc/bulletinc.dat

**Seeing, time resolution and statistical error on single transit timing**

The warm air rising up near the external side of the pinhole produces turbulence. Since the circular pinhole used for the winter solstice 2011 is 1.6 cm of diameter, warm air vortices of such dimension are affecting the entire objective pupil, and therefore the image on the focal plane shakes as a whole.

Moreover the limit imposed by the diffraction is intriguingly similar to the one found by the seeing. For a pinhole of diameter d=1.6 cm and for a light of wavelength λ=550 nm, $\Delta\vartheta$=8.7 arcsec is the angular resolution for point-like sources according to the Rayleigh resolution criterion ($\Delta\vartheta=1.22\cdot\lambda/D$).

In terms of time resolution, through the equation

$$\Delta\vartheta = 15 \text{ arcsec} \cdot \cos\delta \cdot \Delta t$$

where δ is the solar declination at the moment of the observation[3] we obtained a $\Delta\vartheta$=4.5 arcsec which represents the better angular resolution achievable at the Clementine Gnomon under ordinary seeing conditions.

The method chosen for measuring the transit time with an expected accuracy of a few hundredths of second was described in a paper of 2006.[4] The strategy consisted to average the transit times of the preceding and proceeding limb on lines parallel and evenly spaced with respect to the meridian line. N lines and 2N contact times were determined each one with 1/30 s of accuracy through video inspection frame by frame. A final resolution on the averaged value of the central time calculated over the above N lines was expected to $\sqrt{N}$ times the accuracy of a single measurement, so with N=30 lines 0.1 s should be attained.

But no convergence at all occurred having 30 lines and 30 supposedly independent determination of the central time. This was because the 30 determinations or the 30 transits above 30 parallel lines occurred over more than 5 minutes were not statistically independent one from another.

It means that the turbulent phenomena in the local atmosphere which affect the image have timescales which are comparable with the whole duration of the 30 measurements, and not only very rapid behavior ranging around the 20-50 Hz of typical frequency. The contributions of very rapid fluctuations over a long sampling time interval would be randomly averaged to zero.

The measurements of timing of Santa Maria degli Angeli did not attain the expected accuracy because the single timings were not Gaussian-like distributed, so their average did not converge as Gaussian data would do.

Studies conducted in Locarno and Paris with much greater telescopes (45 and 33 cm respectively) confirmed the non-Gaussian nature of the turbulent phenomena of the atmosphere, responsible of a tiny motion of the whole solar image during the time of a drift-scan transit (typically lasting 2 minutes and half).[8]

This fact implies that the average of N durations of consecutive transits does not converge with a Gaussian decreasing statistical dispersion proportional to $1/\sqrt{N}$.[5,7]

Therefore to validate a drift-scan measurement of the solar diameter a measurement of the drift of the solar center made in parallel by another large field (6 to 8 solar radii) instrument would be necessary.

**Position of the actual pinhole**

A simpler measurement to do in the Basilica at the moment of the meridian transit is the one on the position of the solar image with respect to the original marks of 1700.

This operation is complicate by the background illumination of the Basilica, which once (300 years ago) was darkened by tents during daytime observational sessions.

Nevertheless the lower contrast between the image and the background produces an error which is symmetrical for the northern and the southern limb of the Sun. Therefore this error does not affect the determination of the position of the center of the Sun.

The measure has been done as in the following.

Three sheets of papers have been attached to the floor next to the meridian line.

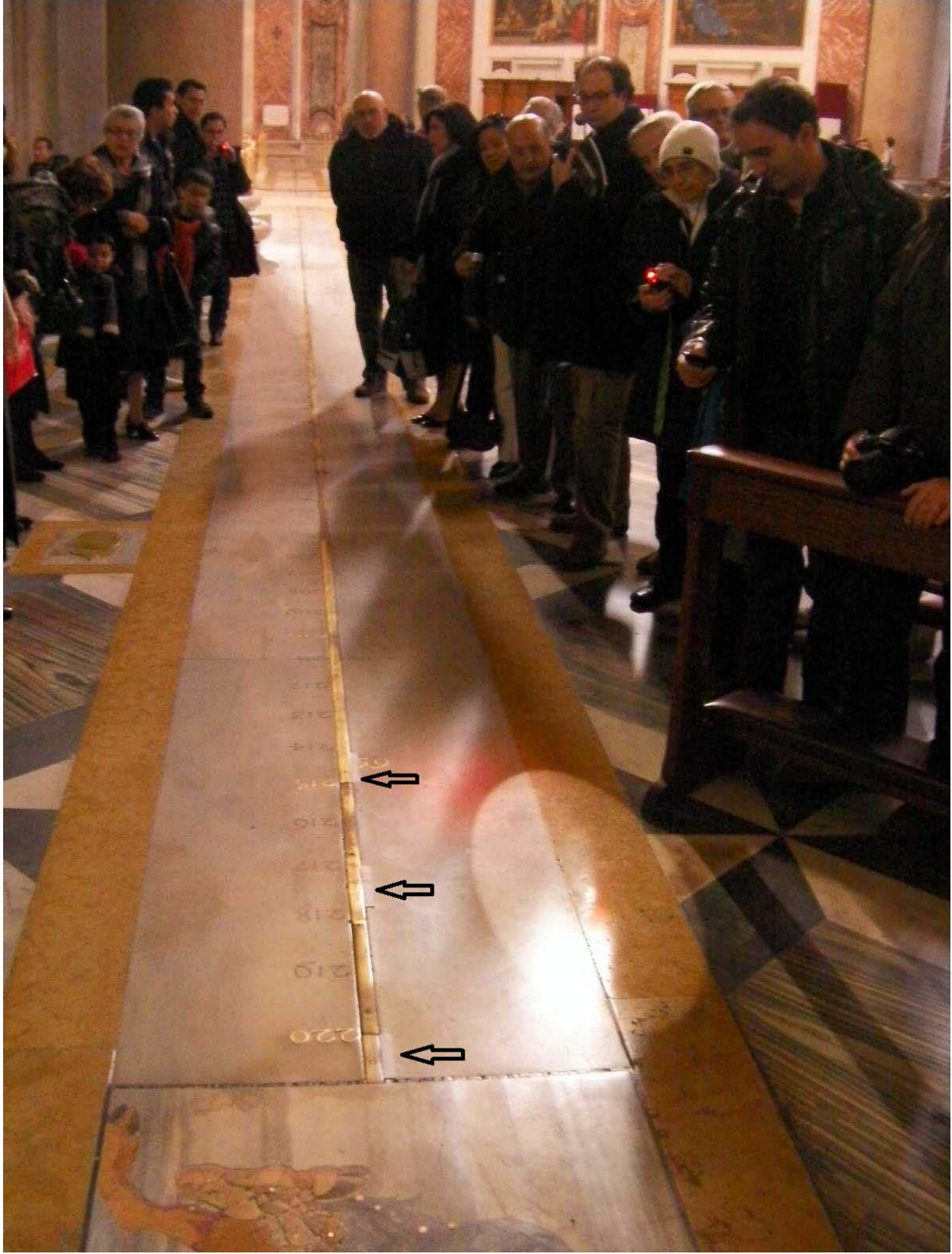

**Fig. 2** Disposition of the white papers next to the meridian line. The approaching image of the Sun of December 21, 2011 is on the west (right) side of the meridian, the photo is taken in the direction of the pinhole.

When the transit occurred with a pencil the upper and the lower limbs were signed over the paper (in the image the papers are on the right side of the meridian line); also the central contact point position was evaluated.

After the position was calculated in terms of the original units of measurements: the 1/100 of the height of the original pinhole (20.344 m).
This unit of measurement allows to have directly the tangent of the zenithal angle of each limb.
The computation of solar ephemerides has been done by using the Ephemvga ephemeris program.[6]

The following tables resume all our measurements.

| Date December 2011 | Transit time Obs. [ ±0.3 s] | Transit time Calc. | O-C [ ±0.3 s] | Pressure hPa | Air Temp °C |
|---|---|---|---|---|---|
| 20 | 12:07:42.0 | 12:07:24.2 | +17.8 | 1016 | 9 |
| 21 | 12:08:12.5 | 12:07:54.1 | +18.4 | 1013 | 11 |
| 22 |  | 12:08:24.0 |  | 1017 | 11 |

Air temperatures and pressure from Roma Urbe airport meteo station.[2]

| Date Dec 2011 | N limb Obs. | S limb Obs. | Center Obs. | Center Calc. | Center O-C mm |
|---|---|---|---|---|---|
| 20 | 214.6657 | 220.0885 | 217.3494 | 217.3754 | -5.29 |
| 21 | 214.7419 | 220.204 | 217.44525 | 217.4746 | -5.97 |
| 22 | 214.7862 | 220.2015 | 217.46615 | 217.49544 | -5.96 |
|  |  |  |  | average | -5.7±0.4 |

The unit of measure of the positions of the limbs is 1=203.44 mm, i.e. 1/100 of the height of the pinhole (as measured on Feb 2$^{nd}$ 2006 [4,11]). Each measurement has been computed from the lines traced on the paper, e.g. on Dec. 21 the Southern limb reached 4.15 mm above the 220 sign, so 220+4.15/203.44=220.204.
The image of the Sun covers 5.43 of such units, having a diameter of 1951 arcsec. Therefore the following equivalence is valid (having taken into account also the angular dimension of the pinhole of 67 arcsec):
1 arcsec = 0.55 mm
The correction of temperature and pressure corresponds to less than 0.2 mm.
The correction to the refraction formula (used at the first order i.e with the formula 60"·tan(z)), proportional to tan³(z), it is about ~0.01 arcsec, i.e. less than 0.005 mm.
The error induced by the contrast, which changed from one day to the other, is assumed as the same on both limbs, and the central position was not affected by this effect.
The error induced by the [18.1±0.3] s of delay with respect to the true meridian transit is because the Sun is slightly lower than its maximum height at that moment, but its position along the line changes for less than 0.08 mm.

---

[2] http://www.meteorologic.net/archives-metar_LIRU.html?jour=20&mois=12&year=2011

To compute the position of the solar center, the differential refraction for the Sun at ~24.7° above the horizon, has been also taken into account.

**Conclusions:**

The difference of 5.7±0.4 mm left between ephemerides and measured values is systematic.
It can be due either to
1. Different position of the center of the pinhole with respect to the 1702 original position.
In this case the pinhole had to be 5.7±0.4 mm northward than nowadays. This can be suggested by a figure published by the astronomer Francesco Bianchini (1703) who designed and realized the instrument.[9,10,11]

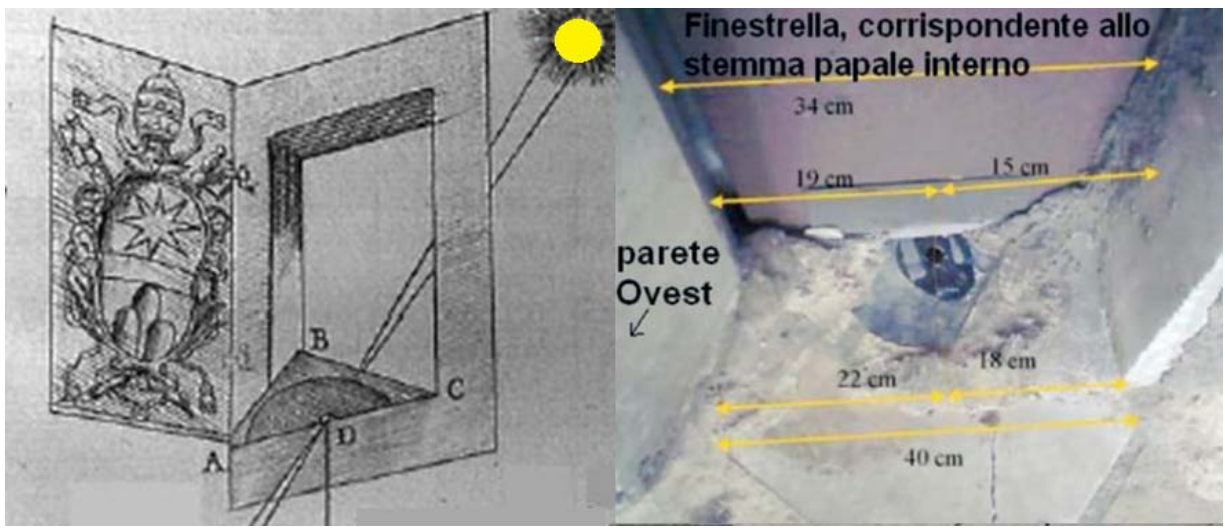

**Fig. 3** Left: the original figure of Bianchini (1703) showing the solar pinhole.
It's center D lays on the window's border AC. Right the actual position of the pinhole, as seen from the external side, is not on the border but more outward.

2. Different height of the pinhole with respect to the original measurement, due to thermal expansion of the concrete of Diocletian Baths. In this case Bianchini could have measured the height of the pinhole in Summer time, when the air temperature was about 10° C higher than in the winter and the pinhole was slightly higher.
Assuming a typical thermal expansion coefficient of $10^{-5}$ /°C for the roman concrete (Opus Caementitium)[3] for a 10 °C increase of temperature from winter to summer, we have for the 20 m wall of Diocletian Baths, where the pinhole is built-in, a raising of 2 mm.
With the inclination of 25° this corresponds to a 4.3 mm shift northward of the ending points of the meridian line. The thermal coefficient of the roman concrete is probably with a 30% uncertainty, and it can be derived by comparison with further summer time measurements.
Probably the height of the pinhole was measured in summer (1701 or 1702) and the marks on the meridian line were made consequently.

---

[3] http://www.romanconcrete.com/ and also http://en.wikipedia.org/wiki/Roman_concrete

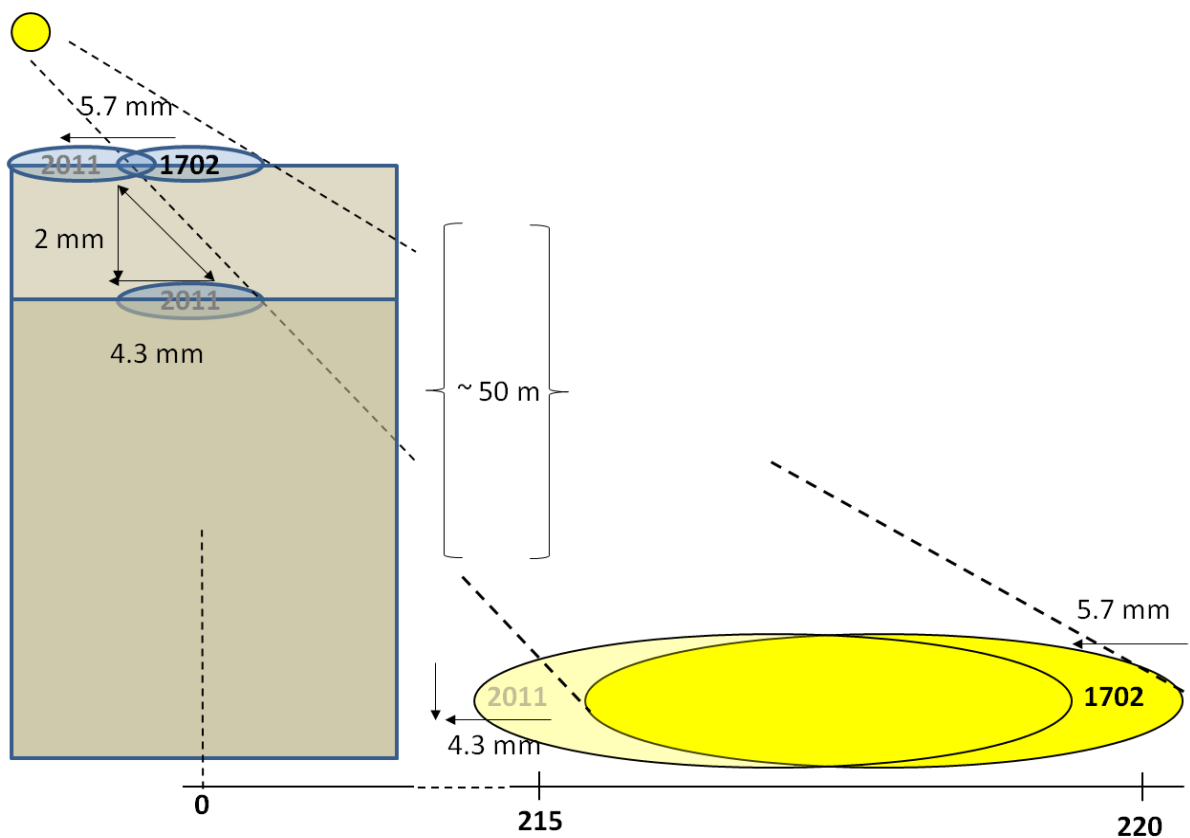

**Fig. 4** A vertical shift of the pinhole position is due to the thermal expansion between winter 2011 and summer 1702 (or 1701), when the height of the pinhole could have been measured. An horizontal shift between 1702 and 2011 position is also possible. Both shifts bring the image backwards of some millimeters. The two effects are probably combined together.

A forthcoming restoration will concern the pinhole, and the window in which it is located, that was opened to observe stellar transits as well.
The position of the pinhole will be chosen in order to reproduce correctly the measurements of the center of the Sun during the winter solstice when the graduation marks near the end point of the meridian line (215-220) are used as references.
The diameter of the pinhole will be restored to the original 20 mm, i.e. 1/1000 of the height of the pinhole.
The platform containing the pinhole will be realized in invar, and colored with white Titanium dioxide.